\documentclass[camera,letterpaper,nomarginnotes,nonarrowgutter]{jpaper}
\usepackage{listings}
\usepackage{fancyhdr}

\usepackage{datetime}

\usepackage{cite}
\usepackage{amsmath,amssymb,amsfonts}
\usepackage{algorithmic}
\usepackage{graphicx}
\usepackage{textcomp}
\usepackage{xcolor}
\usepackage{tikz}

\usepackage[utf8]{inputenc}
\usepackage[T1]{fontenc}

\usepackage{booktabs} % For formal tables
\usepackage{setspace}
\usepackage[italic]{mathastext}
\usepackage{array}
\usepackage{titlesec}
\usepackage[normalem]{ulem}
\usepackage{multirow}
\usepackage{multicol}
\usepackage{color}
\usepackage[font={small,bf}]{caption}
\usepackage{float}
\usepackage[font={small,bf}]{subcaption}
\usepackage[linesnumbered,ruled]{algorithm2e}
\usepackage{makecell}

\usepackage{pifont}
\usepackage[]{microtype}
            
\usepackage{todonotes} % For margin notes
\usepackage{marginnote} 
\usepackage{pifont}
\let\marginpar\marginnote
\usepackage[bookmarks=true,breaklinks=true,colorlinks,linkcolor=black,citecolor=blue,urlcolor=black]{hyperref}
\definecolor{denim}{rgb}{0.08, 0.38, 0.74}
\definecolor{darkolivegreen}{rgb}{0.33, 0.42, 0.18}
\definecolor{dgreen}{rgb}{0.00, 0.75, 0.00}
\definecolor{darkpink}{rgb}{0.88, 0.28, 0.54}
\definecolor{forestgreen}{rgb}{0.0, 0.27, 0.13}
\definecolor{amber}{rgb}{1.0, 0.49, 0.0}
\definecolor{lightyellow}{rgb}{0.980, 0.956, 0.623}
\definecolor{lightblue}{rgb}{0.980, 0.956, 0.623}
\definecolor{darkamber}{rgb}{0.5, 0.19, 0.0}
\definecolor{dkgreen}{rgb}{0,0.6,0}
\definecolor{gray}{rgb}{0.5,0.5,0.5}
\definecolor{mauve}{rgb}{0.58,0,0.82}
\definecolor{lightmauve}{rgb}{0.68,0.4,0.92}
\definecolor{chocolate}{rgb}{0.48, 0.25, 0.0}
\definecolor{dollarbill}{rgb}{0.52,0.73,0.4}
\definecolor{dkdkgreen}{rgb}{0,0.45,0}
\definecolor{gfored}{rgb}{0.580, 0.050, 0.211}
\definecolor{darkwarmgray}{rgb}{0.15, 0.050, 0.05}
\definecolor{ups-truck}{rgb}{0.53, 0.28, 0.21}

\newcommand\rev[1]{{\color{black}{#1}}}
\newcommand\revb[1]{{\color{black}{#1}}}
\newcommand\revc[1]{{\color{black}{#1}}}
\newcommand\revd[1]{{\color{black}{#1}}}
\newcommand\reve[1]{{\color{black}{#1}}}
\newcommand\revf[1]{{\color{black}{#1}}}

\makeatletter
\g@addto@macro{\normalsize}{%
  \setlength{\abovedisplayskip}{2pt plus 1pt minus 1pt}
  \setlength{\belowdisplayskip}{2pt plus 1pt minus 1pt}
  \setlength{\intextsep}{2pt plus 1pt minus 1pt}
  \setlength{\textfloatsep}{3pt plus 1pt minus 1pt}
  \setlength{\dbltextfloatsep}{3pt plus 1pt minus 1pt}
  \setlength{\skip\footins}{4pt plus 1pt minus 1pt}
}
\def\BibTeX{{\rm B\kern-.05em{\sc i\kern-.025em b}\kern-.08em
    T\kern-.1667em\lower.7ex\hbox{E}\kern-.125emX}}

\def\UrlBreaks{\do\/\do-\/\do.\/\do:}

\expandafter\def\expandafter\UrlBreaks\expandafter{\UrlBreaks
  \do\a\do\b\do\c\do\d\do\e\do\f\do\g\do\h\do\i\do\j
  \do\k\do\l\do\m\do\n\do\o\do\p\do\q\do\r\do\s\do\t
  \do\u\do\v\do\w\do\x\do\y\do\z\do\A\do\B\do\C\do\D
  \do\E\do\F\do\G\do\H\do\I\do\J\do\K\do\L\do\M\do\N
  \do\O\do\P\do\Q\do\R\do\S\do\T\do\U\do\V\do\W\do\X
  \do\Y\do\Z}
  
\newcommand{\squishlist}{
 \begin{list}{$\circ$}
  { \setlength{\itemsep}{0pt}
     \setlength{\parsep}{0pt}
     \setlength{\topsep}{0pt}
     \setlength{\partopsep}{0pt}
     \setlength{\leftmargin}{1em}
     \setlength{\labelwidth}{1em}
     \setlength{\labelsep}{0.5em} } }

\newcommand{\squishsublist}{
\begin{list}{$\rightarrow$}
 { \setlength{\itemsep}{0pt}
    \setlength{\parsep}{0pt}
    \setlength{\topsep}{-10em}
    \setlength{\partopsep}{-3pt}
    \setlength{\leftmargin}{1em}
    \setlength{\labelwidth}{1em}
    \setlength{\labelsep}{0.5em} } }

\newcommand{\squishend}{
  \end{list}  }

\usepackage{titlesec}
\titlespacing*{\section}{0pt}{0.3ex}{0.1ex} % Adjust spacing for sections
\titlespacing*{\subsection}{0pt}{0.3ex}{0.3ex} % Adjust spacing for subsections
\titlespacing*{\subsubsection}{0pt}{0.3ex}{0.3ex} % Adjust spacing for subsections

\def\BibTeX{{\rm B\kern-.05em{\sc i\kern-.025em b}\kern-.08em
    T\kern-.1667em\lower.7ex\hbox{E}\kern-.125emX}}
\begin{document}

\title{\emph{Invited:} Accelerating Genome Analysis \\ via Algorithm-Architecture Co-Design}

\newcommand{\affilETH}[0]{\small {$$}}
\author{\vspace{-18pt}\\%
% \fontsize{11}{12}\selectfont%
{Onur Mutlu}\quad%
{Can Firtina}\quad%
\vspace{-3pt}\\%
% {\fontsize{10}{11}\selectfont
% \qquad\qquad\qquad\qquad\qquad\qquad
\affilETH\emph{ETH Z{\"u}rich}%
% \qquad\qquad
% \affilCMU\emph{Carnegie Mellon University}%
% }
\vspace{-12pt}}

\maketitle
\thispagestyle{plain}

% \setstretch{0.82}

\begin{abstract}

High-throughput sequencing (HTS) technologies have revolutionized the field of genomics, enabling rapid and cost-effective genome analysis for various applications. However, the increasing volume of genomic data generated by HTS technologies presents significant challenges for computational techniques to effectively analyze genomes. To address these challenges, several algorithm-architecture co-design works have been proposed, targeting different steps of the genome analysis pipeline. These works explore emerging technologies to provide fast, accurate, and low-power genome analysis.

This paper provides a \revb{brief} review of the recent advancements in accelerating genome analysis, covering the opportunities and challenges associated with the acceleration of the key steps of the genome analysis pipeline. Our analysis highlights the importance of integrating multiple steps of genome analysis using suitable architectures to unlock significant performance improvements and reduce data movement and \revb{energy} consumption.~\revb{We conclude by emphasizing} the need for novel strategies and techniques to address the growing demands of genomic data generation and analysis.

\end{abstract}

\section{Introduction} \label{sec:introduction}

Genome analysis plays a crucial role in various fields such as personalized medicine~\cite{pickar-oliver_next_2019
% , lightbody_review_2019, doudna_promise_2020
}, agriculture~\cite{shahroodi_demeter_2022}, evolutionary biology~\cite{kanehisa_toward_2019}, pharmacogenomics~\cite{morganti_next_2019}, infectious disease control~\cite{
dunn_squigglefilter_2021,
alser_covidhunter_2022},
% forensics~\cite{bruijns_massively_2018},
cancer research~\cite{lawrence_mutational_2013
% , vogelstein_cancer_2013
} and microbiome studies~\cite{johnson_evaluation_2019}. The advent of \rev{high-throughput sequencing (HTS) technologies, such as sequencing-by-synthesis (SBS)~\cite{bentley_accurate_2008}, Single Molecule Real-Time (SMRT)~\cite{eid_real-time_2009}, and nanopore sequencing~\cite{branton_potential_2008, deamer_three_2016, senol_nanopore_2018}}, has revolutionized genome analysis, enabling faster and more cost-effective sequencing of genomes by generating a large amount of genomic data at relatively low cost~\cite{shendure_dna_2017}. However, the analysis of genomic data is challenging \rev{due to a variety of reasons:} 1)~HTS technologies can \revb{only} sequence \rev{relatively} short fragments of genomes, called \emph{reads}, whose \rev{locations in the entire genome} are unknown, 2)~\rev{these} reads can contain \emph{sequencing errors}~\cite{shendure_dna_2017, firtina_apollo_2020}, leading to differences from their original sequences, \rev{3)~the sequenced genome may not (and usually does not) exactly match recorded genomes in a reference database, known as \emph{reference genomes}, due to variations between individuals within and across species}. Despite significant improvements in computational tools since the 1980s~\cite{alser_technology_2021} to overcome \rev{such} challenges, the rapid growth in \rev{genomic} \rev{data~\cite{stephens_big_2015}} has led to \rev{ever larger} computational overhead\rev{s} in the genome analysis pipeline, posing \rev{large} challenges for efficient and timely analysis of genomes~\cite{alser_accelerating_2020, alser_molecules_2022}.
    
% To accurately and quickly analyze genomes, several computationally costly steps are taken in the genome analysis pipeline, which also impacts the energy consumption of the process.
\rev{A genome analysis pipeline consists of multiple key steps, each of which affects the accuracy, speed, and energy consumption of genome analysis.}
First, \emph{basecalling} translates the \emph{raw sequencing data} that HTS generates \rev{(e.g., measured electrical signals in nanopore sequencing)} into sequences of genomic characters (e.g., A, \revd{C}, G, and \revd{T}s in DNA). Basecalling is \rev{time-consuming} because it relies \rev{heavily} on \revb{compute-intensive} approaches that process large chunks of noisy and error-prone raw data to accurately infer the \revd{actual} nucleotide sequences~\cite{purnell_nucleotide_2008, timp_dna_2012, boza_deepnano_2017, senol_nanopore_2018, rang_squiggle_2018, xu2021fast, alser_molecules_2022}.
Second, \rev{\emph{real-time analysis of raw sequencing data}~\cite{loose_real-time_2016, edwards_real-time_2019, kovaka_targeted_2021, zhang_real-time_2021, payne_readfish_2021, dunn_squigglefilter_2021, bao_squigglenet_2021, shih_efficient_2022, sadasivan_rapid_2023, senanayake_deepselectnet_2023, firtina_rawhash_2023}} aims to analyze the reads simultaneously while the read is being sequenced using a particular sequencing technology (e.g., nanopore sequencing). Although real-time analysis of raw sequencing data provides enormous advantages in significantly reducing the overall genome analysis time and cost~\cite{loose_real-time_2016}, it introduces unique challenges as the analysis 
% is required to be as fast as the throughput of sequencers 
\rev{needs to match \revb{stringent} throughput and latency requirements}
to \rev{satisfy \emph{real-time} requirements}~\cite{firtina_rawhash_2023}.
\rev{Third, \emph{read mapping} aims to find similarities and differences between genomic sequences (e.g., between sequenced reads and reference genome\revb{s} of one or more species).}
% Third, \emph{read mapping} aims to find similarities and differences between genomic sequences, which is useful for reconstructing their corresponding origin information lost during sequencing (e.g., genomic positions of reads).
Read mapping includes several steps such as sketching~\cite{schleimer_winnowing_2003, roberts_reducing_2004, li_minimap_2016, firtina_blend_2023, baker_dashing_2023, joudaki_aligning_2023}, seeding~\cite{altschul_basic_1990, langmead_ultrafast_2009, xin_accelerating_2013, xin_shifted_2015, xin_optimal_2016, alser_gatekeeper_2017, li_minimap2_2018, alser_shouji_2019, alser_sneakysnake_2020}, and alignment~\cite{needleman_general_1970, smith_identification_1981, baeza-yates_new_1992, myers_fast_1999, senol_cali_genasm_2020, marco-sola_fast_2021}, which demand considerable processing power and memory due to the large scale of genomic sequences~\cite{alser_technology_2021, kim_airlift_2021, kim_fastremap_2022}.
\rev{Fourth, subsequent steps of the genome analysis (i.e., \emph{downstream analysis}) use the output generated in the read mapping step. An example of such downstream analysis is known as
% include 1)~constructing \emph{de novo} genome assembly~\cite{senol_cali_genasm_2020}, metagenomics~\cite{wood_improved_2019, lapierre_metalign_2020}, and
\emph{variant calling}~\cite{kwok_comparative_1994, nickerson_polyphred_1997, marth_general_1999, weckx_novosnp_2005, li_mapping_2008, poplin_scaling_2018, poplin_universal_2018},} which
% Fourth, the output generated in the read mapping step can be used in various applications such as \emph{de novo} genome assembly~\cite{senol_cali_genasm_2020}, metagenomics~\cite{lapierre_metalign_2020, wood_improved_2019}, and variant calling~\cite{poplin_scaling_2018, poplin_universal_2018}, in the \emph{downstream analysis}.
% Downstream analysis often requires additional computationally intensive steps~\cite{alser_molecules_2022}, including
% set intersection~\cite{liu_cmash_2022}, graph processing~\cite{li_minimap_2016, firtina_aphmm_2022}, or
% the use of deep neural networks \rev{(DNNs)}~\cite{poplin_universal_2018}.
% These additional steps further contribute to the overall computational overhead and energy consumption of the genome analysis pipeline~\cite{alser_molecules_2022}.
% Fifth, \emph{variant calling}
aims to identify \revb{genetic \revc{differences}, known as \emph{variants},} between an individual's genome and a reference genome. Variant calling is often followed by additional steps, such as \emph{gene annotation}~\cite{guigo_prediction_1992, borodovsky_detection_1995, burge_prediction_1997, li_snap_2007, wang_annovar_2010} and \emph{enrichment analysis}~\cite{ashburner_gene_2000, doniger_mappfinder_2003, subramanian_gene_2005, otlu_glanet_2017}. These steps aim to
generate insights from the identified variants and determine if these variants show
% assign the functional information (e.g., genes) to the identified variants and determine if the functional information shows
an unexpectedly high or low statistical \revb{correlation} with specific functional \revb{behavior} (e.g., association with a disease) that can be used in a clinical report~\cite{biesecker_diagnostic_2014}.

Many pure algorithmic and software techniques aim to address the computational challenges in the genome analysis pipeline. These works improve the \rev{performance and accuracy} of the computational tools by 1)~reducing overall computational and space complexity~\cite{marco-sola_fast_2021, marco-sola_optimal_2023}, 2)~eliminating useless work~\cite{xin_accelerating_2013, xin_shifted_2015, xin_optimal_2016, bingol_gatekeeper-gpu_2021, kim_airlift_2021, kim_fastremap_2022, cavlak_targetcall_2022, singh_framework_2022, firtina_blend_2023}, 3)~optimizing data structures and memory access patterns~\cite{hach_mrsfast_2010, pan_kmerind_2016, ellis_dibella_2019}, 4)~exploiting parallelism in multi-core, many-core, \rev{and SIMD} architectures~\cite{xin_shifted_2015, daily_parasail_2016, vasimuddin_efficient_2019, cavlak_targetcall_2022, singh_framework_2022, singh_fpga-based_2021, kalikar_accelerating_2022, firtina_blend_2023, lindegger_scrooge_2023}, and 5)~employing machine learning techniques~\cite{poplin_universal_2018, firtina_apollo_2020, cavlak_targetcall_2022, singh_framework_2022}.
% , and adapting algorithms to take advantage of hardware-specific features, such as SIMD instructions~\cite{}.
\rev{These works fall short on greatly improving performance and energy consumption due to at least three major reasons.}
% However, these pure software designs have fundamental limitations in enhancing the performance of these computational tools while also demanding high power usage for several reasons.
First, many of these \rev{approaches} incur significant data movement \rev{between} computation units and memory \revb{units}~\cite{alser_accelerating_2020, mansouri_ghiasi_genstore_2022}.
\rev{Second, a large portion of the data becomes useless in downstream genome analysis~\cite{mao_genpip_2022}, \revb{and performing computation on it} wastes \reve{time} and energy.}
\rev{Third, HTS technologies produce sequencing data at an increasingly high rate, which \revb{makes it} challenging to keep up with the throughput of these sequencing technologies, especially in time-critical scenarios~\cite{alser_accelerating_2020, firtina_rawhash_2023}.}

\rev{Since software techniques alone are not effective \revd{enough} at coping with huge amounts of genomic data and \revb{the} stringent requirements of genome analysis, it is critical to design software-hardware cooperative techniques to accelerate genome analysis. To this end, several works co-design algorithms and architectures to substantially improve the performance and \revd{energy efficiency} of the genome analysis pipeline. These works 1)~reduce data movement overhead\revd{s} by \revd{employing} processing in memory (PIM)~\cite{kaplan2017resistive, kim_grim-filter_2018, 
kaplan2018rassa,
% lou2018brawl, zokaee2018aligner, huangfu_radar_2018,
gupta2019rapid,
% zokaee2019finder,
angizi2019aligns, ghose_processing--memory_2019-1,
angizi2020exploring,
lou2020helix, chen2020parc,
chowdhury_dna_2020,
angizi2020pim,
kaplan2020bioseal, laguna2020seed, khatamifard2021genvom, khalifa_filtpim_2021,
li2021pim,
diab_high-throughput_2022, diab_framework_2022, shahroodi_demeter_2022}, \revd{or} processing near storage (e.g., solid-state drives)~\cite{mansouri_ghiasi_genstore_2022} and 2)~efficiently \revd{co-design} and execute computationally complex algorithms with massive parallelism and efficient hardware design using \revb{specialized architectures}, e.g., field programmable gate arrays (FPGAs) and application-specific integrated circuits (ASICs)~\cite{madhavan2014race,
chen2014accelerating,
waidyasooriya2015hardware, chen2015novel, chen2016spark, goyal2017ultra,
alser_gatekeeper_2017,
% alser_magnet_2017,
banerjee2018asap, rucci2018swifold,
fei2018fpgasw, fujiki2018genax, turakhia2018darwin,
% wu2019fpga,
alser_shouji_2019, alser_sneakysnake_2020, fujiki2020seedex, wu_fpga-accelerated_2020, senol_cali_genasm_2020,
% li2021pipebsw, chen2021high,
haghi_fpga_2021, singh_fpga-based_2021,
% hammad_scalable_2021, ramachandra_ont-x_2021,
dunn2021squigglefilter, shih_efficient_2022, firtina_aphmm_2022,
% wu_fpga_2022,
senol_cali_segram_2022}}

% The exploration of algorithm-architecture co-design for accelerating genome analysis holds great potential in enhancing the capabilities of various genomics applications by better understanding and adapting to emerging hardware technologies and addressing the computational bottlenecks in recent computational tools.
%, including personalized medicine, agriculture, and evolutionary biology, by enabling faster and more efficient analysis of genomic data.

In this paper \rev{(and the associated invited talk)}, we review the recent advancements in accelerating genome analysis via algorithm-architecture co-design and discuss emerging challenges that highlight the need for new acceleration techniques. We aim to provide a \rev{brief yet} comprehensive overview of the current state of the field and inspire future research directions to further improve the efficiency of genome analysis \revb{and hopefully enable new use cases and computing platforms}.

\section{Accelerating Basecalling} \label{sec:basecalling}

HTS technologies produce raw sequencing data, the content of which depends on the type of sequencing technology employed. There are three main types of sequencing technologies: sequencing by synthesis (SBS)~\cite{bentley_accurate_2008}, Single Molecule Real-Time (SMRT)~\cite{eid_real-time_2009}, and nanopore sequencing~\cite{branton_potential_2008}. SBS generates \rev{images where the color intensity at a particular position of an image represents the base of the read.} Basecalling after SBS aims to accurately associate these colors with their corresponding bases \rev{while correcting sequencing errors}~\cite{cacho_comparison_2016}. SMRT sequencing generates continuous images in a movie format by sequencing the same read multiple times \revb{via a} strategy known as circular consensus sequencing (CCS)~\cite{wenger_accurate_2019}. Although these images can be \revb{quickly} converted to their corresponding bases, the high noise associated with SMRT sequencing requires additional steps to correct sequencing errors~\cite{wenger_accurate_2019}. These techniques include alignment~\cite{li_minimap2_2018}, consensus assembly construction~\cite{wenger_accurate_2019}, and polishing~\cite{chin_nonhybrid_2013, firtina_apollo_2020}. Nanopore sequencing generates raw electrical signals as DNA or RNA molecules pass through tiny pores (i.e., nanoscale holes) called \emph{nanopores}~\cite{branton_potential_2008}. Changes in ionic current, measured as nucleotides pass through, are sampled in real-time and used to perform 1)~basecalling and 2)~real-time genome analysis.

Recent basecalling works~\cite{
boza_deepnano_2017, lv_end--end_2020,
zeng_causalcall_2020, xu2021fast,
konishi_halcyon_2021, perevsini2021nanopore, yeh_msrcall_2022,
huang_sacall_2022, singh_framework_2022, cavlak_targetcall_2022} \revb{especially} focus on basecalling raw nanopore signals due to two \rev{major} reasons. First, the measured signal represents a combination of \emph{multiple nucleotides} passing through the nanopore, making the basecalling task more challenging compared to the relatively simpler and more direct signal-to-base conversion in SBS and SMRT sequencing methods~\cite{alser_molecules_2022, singh_framework_2022}. Second, nanopore sequencing provides \rev{unique} opportunities for real-time genome analysis \rev{that can be used to reduce the time and cost of \revc{sequence analysis}~\cite{alser_molecules_2022, firtina_rawhash_2023}, as we discuss in \revb{\S}\ref{sec:realtime}.}

\revb{B}asecalling techniques developed for nanopore sequencing mainly use deep neural networks (DNNs)~\cite{singh_framework_2022} to achieve high accuracy. However, these methods are computationally expensive \rev{to train and use} with large amount\revc{s} of raw electrical signal data~\cite{mao_genpip_2022}. To address this issue, several algorithm-architecture co-design works have been proposed.
First, \revb{some works} accelerate the execution of DNN operations using graphics processing units (GPUs)~\cite{
boza_deepnano_2017, lv_end--end_2020,
zeng_causalcall_2020, xu2021fast,
konishi_halcyon_2021, perevsini2021nanopore, yeh_msrcall_2022,
huang_sacall_2022}. GPUs can substantially improve basecaller performance by providing massive parallelism for performing matrix multiplications \rev{in DNNs}.
Second, RUBICON~\cite{singh_framework_2022} and TargetCall~\cite{cavlak_targetcall_2022} \rev{reduce} \revc{unnecessary} computations in GPU-based basecallers by
% employing less complex DNN architectures with high accuracy. These works demonstrate that \rev{DNN architectures can be optimized} to improve performance by
\rev{1)~reducing the DNN parameters and precision~\cite{singh_framework_2022} or 2)~introducing pre-basecalling filters~\cite{cavlak_targetcall_2022}.}
\rev{Third, several works use processing-in-memory (PIM)~\cite{lou2018brawl, lou2020helix, mao_genpip_2022}, or FPGAs~\cite{wu_fpga-accelerated_2020
% ramachandra_ont-x_2021, hammad_scalable_2021, wu_fpga_2022
} to accelerate basecalling and reduce power consumption. A recent work that uses PIM, GenPIP~\cite{mao_genpip_2022}, 
% by minimizing large data movement in basecallers and providing efficient execution of computationally costly steps.
% Although PIM can reduce data movement overhead, GenPIP~\cite{}
shows that a significant portion of \revb{useless} data can propagate to downstream analysis, causing \revc{unnecessary} data movement, compute cycles, and energy consumption. To eliminate \revb{such useless} operations, GenPIP combines \emph{both} basecalling and read mapping in PIM to quickly identify \revb{unnecessary} reads without fully basecalling them, \revb{thereby} reducing \emph{both} data movement overhead\revc{s} and overall \reve{execution time} \revc{spent} in basecalling and read mapping.}

We believe that integrating multiple steps of genome analysis using suitable architectures, such as PIM, can unlock significant opportunities for \rev{1)~reducing data movement overhead\revb{s}, 2)~eliminating useless basecalling, and 3)~avoiding useless data movement and computation in downstream analysis. These approaches have the potential to substantially enhance the performance and energy efficiency of the entire genome analysis pipeline.}
\section{Accelerating Real-Time Genome Analysis} \label{sec:realtime}

Real-time genome analysis aims to perform the steps in the genome analysis pipeline (e.g., read mapping) while the raw sequencing data \revb{is} generated~\cite{loose_real-time_2016, firtina_rawhash_2023}. The main challenges of real-time genome analysis are to 1)~match the throughput at which the raw sequencing data is generated, 2)~tolerate the noise in the raw sequencing data to provide accurate \revd{results}, \rev{and 3)~meet the latency and \revd{energy} consumption requirements of \revb{target} applications.} Among the HTS technologies, nanopore sequencing is uniquely suited for real-time genome analysis due to its ability \rev{to eject reads from nanopores without fully sequencing them, known as \emph{adaptive sampling \revb{or Read Until}}~\cite{loose_real-time_2016}.} This feature can significantly reduce the overall sequencing time and cost and reduce the latency of genome analysis by 1)~avoiding full sequencing \revb{of} reads \revb{that will be useless in} downstream analysis and 2)~overlapping the latency of sequencing with steps in downstream analysis.

To enable real-time genome analysis, several works propose \rev{pure algorithmic techniques} or algorithm-hardware co-design solutions. First, ReadFish~\cite{payne_readfish_2021}, ReadBouncer~\cite{ulrich_readbouncer_2022}, and RUBRIC~\cite{edwards_real-time_2019} \revb{use costly basecalling mechanisms for adaptive sampling}. These techniques require \revb{costly and energy-hungry} computational resources. \rev{Such a requirement \revb{may} cause practical challenges in \revb{1)~scaling genome analysis \revc{to} lower energy and cost \revc{levels} and 2)}~performing \revb{in-the-field} sequencing using \revb{mobile} sequencing devices such as ONT MinION~\cite{firtina_rawhash_2023}.} Second, many works such as UNCALLED~\cite{kovaka_targeted_2021}, Sigmap~\cite{zhang_real-time_2021}, and RawHash~\cite{firtina_rawhash_2023} use \rev{efficient techniques to utilize adaptive sampling in low-power devices with usually lower accuracy than the \revd{basecalling mechanisms}. Among these works, RawHash can provide high accuracy for large genomes with \revb{an} efficient and accurate hash-based similarity identification technique.} Third, several algorithm-architecture \revf{co-designs} use FPGAs~\cite{shih_efficient_2022} or ASICs~\cite{dunn2021squigglefilter} to provide fast, accurate, and low-power real-time genome analysis. However, these works are applicable only to small genomes, such as viral genomes, as their algorithm design\revb{s} lack efficient scalability to larger genomes.

We believe that achieving accurate and real-time genome analysis still requires substantial developments in both efficient algorithms and architecture. \rev{This can be achieved by 1)~designing efficient software that can be used in low-power devices for adaptive sampling and real-time genome analysis, 2)~\revb{new techniques for} genome \revb{analysis that do not require} translating the raw sequencing data to \revc{nucleotide} bases, and 3)~\revc{combining} \revb{and parallelizing} several steps in real-time genome \revb{analysis} using efficient algorithm-architecture co-design\revb{s} to minimize the latency \revb{(and energy)} of time-critical genomics applications.}
\section{Accelerating Read Mapping} \label{sec:readmapping}

The goal of read mapping is to identify similarities and differences between genomic sequences, such as between a read and a representative sequence of a species, known as a \emph{reference genome}. Due to \revb{genomic variants} and sequencing errors, \revb{differences and similarities between these sequences (i.e., matches, substitutions, insertions, and deletions) are identified using an approximate string matching (ASM) algorithm to generate an \emph{alignment score} that \revd{quantifies} the degree of similarity between a pair of sequences. \revd{This process is} known as \emph{sequence alignment}. A pair of sequences is \revd{said to be} \emph{aligned} when their alignment score shows a sufficiently high degree of similarity.}
% To efficiently and accurately identify these similarities, read mapping typically involves 1)~finding matching short subsequences between these genomic sequences and 2)~performing approximate string matching (ASM), known as \emph{sequence alignment}, within regions where matching short subsequences are identified.
% Since sequence alignment is computationally costly~\cite{}, performing an alignment after matching subsequences can facilitate a quick and accurate similarity search between a large number of reads and a large reference genome (e.g., three billion bases for a human genome).
However, ASM algorithms often have quadratic time and space complexity, making them computationally challenging for \revd{both long} genomic sequences \revd{and a large number of sequence pairs}. To \revd{ease the identification of} similarities within vast amounts of sequencing data, \revb{read mapping includes multiple steps, such as:} 1)~sketching~\cite{schleimer_winnowing_2003, roberts_reducing_2004, li_minimap_2016, firtina_blend_2023, baker_dashing_2023, joudaki_aligning_2023}, 2)~\revb{indexing} and seeding~\cite{altschul_basic_1990, langmead_ultrafast_2009, xin_accelerating_2013, xin_shifted_2015, xin_optimal_2016, li_minimap2_2018}, 3)~\revb{pre-alignment filtering}~\cite{alser_gatekeeper_2017, alser_magnet_2017, kim_grim-filter_2018, alser_shouji_2019,
alser_sneakysnake_2020, bingol_gatekeeper-gpu_2021}, and 4)~sequence alignment (i.e., ASM)~\cite{needleman_general_1970, smith_identification_1981, baeza-yates_new_1992, myers_fast_1999, senol_cali_genasm_2020, marco-sola_fast_2021}.

Since read mapping is a crucial and computationally expensive step in many genome analysis pipelines, numerous works focus on accelerating it in various ways. First, a significant \reve{fraction} of sequence pairs do \revd{\emph{not}} align, \revb{which leads} to wasted \revb{computation} and energy \revc{during alignment}~\cite{kim_grim-filter_2018}. To avoid \revb{this useless computation}, several works propose \emph{pre-alignment filtering}, \revb{another step in read mapping} that can efficiently detect and \revb{eliminate} highly dissimilar sequence pairs \revb{\emph{without}} using alignment. Most pre-alignment filtering works~\cite{alser_gatekeeper_2017, alser_magnet_2017, kim_grim-filter_2018, alser_shouji_2019, alser_sneakysnake_2020, bingol_gatekeeper-gpu_2021} provide algorithm-architecture co-design using FPGAs, GPUs, and PIM to substantially accelerate the entire read mapping process by exploiting massive parallelism, efficient bitwise operations\revc{,} \revb{and specialized hardware logic} for detecting similarities among a large number of sequences.

% Second, GenStore~\cite{mansouri_ghiasi_genstore_2022} observes that a large portion of sequencing data unnecessarily moves from disk to memory during read mapping, which significantly increases the latency and power consumption in read mapping. To eliminate this redundant data movement, GenStore quickly identifies two primary sets of reads: 1)~reads that are unlikely to align due to high dissimilarity with the reference genome, and 2)~reads that align by exactly matching the reference genome. These two sets of reads are identified within solid-state disks (SSDs) without moving them to the main memory, thus eliminating redundant data movement in the genome analysis pipeline.

Second, GenStore~\cite{mansouri_ghiasi_genstore_2022} observes that a large \reve{amount} of sequencing data unnecessarily moves from \revb{the solid-state drive (SSD)} to memory during read mapping, significantly increasing latency and \revd{energy} consumption. To eliminate this \revb{wasteful} data movement, GenStore \reve{uses specialized logic \emph{within} the SSD to identify} two sets of reads: 1)~reads \revc{that \revf{do} not} align due to high dissimilarity with the reference genome, and 2)~reads that align by exactly matching the reference genome. \reve{Such} reads are \reve{processed} \revb{in the storage system and not moved to main memory or the CPU, thereby} eliminating \revb{unnecessary} data movement in the \revb{system}.

% Third, numerous studies, including GenASM~\cite{senol_cali_genasm_2020} and Darwin~\cite{turakhia2018darwin},
% % and Scrooge~\cite{lindegger_scrooge_2023},
% focus on accelerating the underlying ASM algorithm employed in the sequence alignment
% % (e.g., the Bitap algorithm~\cite{baeza-yates_new_1992})
% through efficient algorithm-architecture co-design utilizing systolic arrays~\cite{fei2018fpgasw}, GPUs~\cite{lindegger_scrooge_2023}, FPGAs~\cite{fei2018fpgasw, fujiki2020seedex, haghi_fpga_2021}, ASICs~\cite{fujiki2018genax}, high-bandwidth memory (HBM)~\cite{senol_cali_segram_2022}, and PIM~\cite{kaplan2017resistive,
% % zokaee2018aligner, huangfu_radar_2018, gupta2019rapid, angizi2019aligns, angizi2020exploring,
% chen2020parc,
% % chowdhury_dna_2020, angizi2020pim, kaplan2020bioseal,
% diab_high-throughput_2022, diab_framework_2022}. These works can achieve substantial speedups of up to several orders of magnitude compared to their CPU baseline, as certain ASM algorithms offer massive parallelism opportunities and efficient bitwise computations
% % that can be effectively exploited with optimized hardware design.
% Among these works, SeGraM~\cite{senol_cali_segram_2022} is the \emph{first} to accelerate aligning sequences to 1)~sequences and 2)~a graph structure that is mainly constructed to represent a large population as a reference genome to reduce the population bias and improve the accuracy of genome analysis.

Third, numerous studies, including GenASM~\cite{senol_cali_genasm_2020} and Darwin~\cite{turakhia2018darwin}, focus on accelerating the underlying ASM algorithm employed in sequence alignment through efficient algorithm-architecture co-design.~\revf{They do so by exploiting} systolic arrays~\cite{fei2018fpgasw}, GPUs~\cite{lindegger_scrooge_2023}, FPGAs~\cite{fei2018fpgasw, fujiki2020seedex, haghi_fpga_2021}, ASICs~\cite{fujiki2018genax}, high-bandwidth memory (HBM)~\cite{senol_cali_segram_2022}, and PIM~\cite{kaplan2017resistive, chen2020parc, diab_high-throughput_2022, diab_framework_2022}. These works \revd{provide} substantial speedups of up to several orders of magnitude compared to \revb{software} baselines. Among these works, SeGraM~\cite{senol_cali_segram_2022} is the \emph{first} to accelerate aligning \revb{sequences to \revb{graphs} that are used to reduce population bias and improve genome analysis accuracy by representing a large population (instead of a few individuals) within a single reference genome.}

% Fourth, there is ongoing research to develop ASM algorithms that further reduce the complexity of overall sequence alignment while still providing either optimal or non-optimal alignment results. For instance, the recent wavefront algorithm (WFA)~\cite{marco-sola_fast_2021} can decrease the quadratic time and space requirements of traditional sequence alignment methods (e.g., the Smith-Waterman algorithm~\cite{smith_identification_1981}), while the more recent bidirectional WFA (BiWFA)~\cite{marco-sola_optimal_2023} achieves linear memory usage compared to the quadratic memory space required by WFA. Although several recent works have accelerated WFA using GPUs~\cite{aguado-puig_wfa-gpu_2023}, FPGAs~\cite{haghi_fpga_2021}, and PIM~\cite{diab_high-throughput_2022}, more optimized algorithm designs like BiWFA hold significant potential in accelerating read mapping using algorithm-architecture co-design, as BiWFA can efficiently scale massively parallel architectures with its linear memory space utilization.

Despite recent advancements, read mapping remains a computational bottleneck in genome analysis~\cite{alser_accelerating_2020, alser_molecules_2022}. This is primarily due to the vast amount of sequencing data generated at an ever-increasing rate by sequencing machines, which puts significant pressure on the mapping step
due to numerous \revc{unnecessary} calculations between dissimilar pairs of sequences.
% Swiftly
% eliminating dissimilar sequences by
Avoiding \revb{wasteful} 1)~data movement, 2)~computation, and 3)~memory space usage \revb{using efficient algorithm-architecture co-design} is \revb{critical} for minimizing the high energy, time, and \revf{storage} costs associated with read mapping and the entire genome analysis pipeline.
\section{Accelerating Variant Calling} \label{sec:variantcalling}
% The objective of variant calling is to identify genetic differences between an individual's genome and a reference genome. These differences, known as variants, are mainly categorized as single-nucleotide polymorphisms (SNPs), insertions, deletions, and larger structural variations (SVs). Accurately and efficiently detecting these variants is crucial for understanding the genetic basis of diseases~\cite{lawrence_mutational_2013}, population genetics~\cite{poplin_scaling_2018}, and evolutionary studies~\cite{kanehisa_toward_2019}.

% The variant calling process typically involves 1)~processing the read mapping output and 2)~variant detection. First, the output of read mapping is usually processed by sorting and optionally identifying the duplicate information in read mapping to minimize the bias introduced when preparing the genomic sample before sequencing in a step called \emph{polymerase chain reaction} (PCR). Removing these duplicates is important to avoid bias that can over- or under-represent the variations~\cite{zverinova_variant_2022}. Second, the mapped reads are analyzed to distinguish genuine variants from sequencing errors or inaccurate read mapping (i.e., misalignment) using costly statistical models~\cite{poplin_scaling_2018} and machine learning techniques~\cite{poplin_universal_2018}.

The objective of variant calling is to identify \revb{genomic variants} between an individual's genome and a reference genome~\cite{kwok_comparative_1994, nickerson_polyphred_1997, marth_general_1999, weckx_novosnp_2005, li_mapping_2008, poplin_scaling_2018, poplin_universal_2018}. These \revb{variants} are mainly categorized as single-nucleotide polymorphisms (SNPs), insertions, deletions, and larger structural variations (SVs). Accurate and efficient detection of these variants is vital for understanding \revb{of} the genetic basis of diseases~\cite{lawrence_mutational_2013}, population genetics~\cite{poplin_scaling_2018}, evolutionary studies~\cite{kanehisa_toward_2019}, personalized medicine~\cite{dong_genome-wide_2019} and pharmacogenomics~\cite{sangkuhl_pharmacogenomics_2020}.

Variant calling involves processing the read mapping output and detecting variants. First, read mapping output is processed by sorting and optionally identifying duplicate information to minimize bias introduced during the \emph{polymerase chain reaction} (PCR) step of sample preparation~\cite{zverinova_variant_2022}. Second, mapped reads are analyzed to distinguish genuine variants from sequencing errors or misalignments using resource-intensive statistical \revb{techniques}~\cite{nickerson_polyphred_1997, weckx_novosnp_2005, poplin_scaling_2018} \revb{or} machine learning techniques~\cite{poplin_universal_2018}.

Variant callers like GATK HaplotypeCaller~\cite{poplin_scaling_2018} \revb{use} costly probabilistic calculations to analyze the likelihood of specific variants in large sequencing datasets. DeepVariant~\cite{poplin_universal_2018}, a DNN-based variant caller, processes read alignment information as images, demanding substantial GPU resources and memory. Reducing computational \revf{requirements} through algorithmic optimizations, parallelization, and efficient data representation is crucial for faster, more accurate genetic variant analyses.

To accelerate variant calling, several \revb{works \revf{propose}} algorithm-architecture \revf{co-designs}. These include fast execution of Pair Hidden Markov Models (Pair HMMs) in FPGAs or ASICs~\cite{ren_fpga_2015, wu_high-throughput_2021}, reducing data movement overhead\revf{s} in GPUs~\cite{li_improved_2021}, and pipelining processing steps with tools like elPrep~\cite{herzeel_multithreaded_2021} and system-on-chip designs~\cite{wu_975-mw_2021}.

Although several works focus on accelerating variant calling, there is an urgent need for \revc{further} \revb{acceleration}, e.g., for DNN-based variant callers that can provide highly accurate results while bypassing certain processing steps, potentially accelerating the entire genome analysis pipeline.

\subsection{Analysis of Variants} \label{subsec:variantcallinganalysis}

% After variant calling, it is essential to analyze the identified variants to comprehend their functional impact on the organism and their role in diseases, population genetics, or evolution. This analysis typically consists of two main steps: gene annotation~\cite{wang_annovar_2010} and enrichment analysis~\cite{otlu_glanet_2017}. Gene annotation adds relevant information to the identified variants, such as genomic locations, associated genes, and predicted functional impacts, using annotation tools and public databases. Enrichment analysis tools identify over- or under-representation of biological processes, molecular functions, or cellular components associated with the annotated variants, offering insights into how these variants may influence specific biological processes or functions. While these tools effectively handle large amounts of data by establishing statistical connections between the identified variants and extensive public datasets, there are, to our knowledge, no recent works focusing on accelerating these steps in the genome analysis pipeline. We believe there is significant potential in integrating these steps with the variant calling process, as the identified variants can immediately be used in genome annotation and enrichment steps without fully generating the entire variant calling information.

Following variant calling, it is \revb{critical} to analyze \revf{the} identified variants to understand their functional impact on the organism and their role in diseases, population genetics, or evolution. This analysis involves gene annotation~\cite{guigo_prediction_1992, borodovsky_detection_1995, burge_prediction_1997, li_snap_2007, wang_annovar_2010} and enrichment analysis~\cite{ashburner_gene_2000, doniger_mappfinder_2003, subramanian_gene_2005, otlu_glanet_2017}. Gene annotation provides relevant information about variants, while enrichment analysis tools identify associations with biological processes, molecular functions, or cellular components. \revc{Although} these tools \revc{need to} \revc{handle} large volumes \revc{of data}, there \revb{is}, to our knowledge, \revb{little} work on accelerating these steps in the genome analysis pipeline. We \revb{believe these steps are critical for acceleration using hardware-software co-design.}
\section{Conclusion and Future Outlook} \label{sec:conclusion}

Rapid advancements in genomic sequencing technologies have led to an exponential increase in generated genomic data. As data generation continues to grow, data movement bottlenecks will increasingly impact performance \revb{and waste energy}~\cite{oliveira_damov_2021, mutlu_modern_2023}. Future research in genome analysis acceleration should focus on \revd{at least} three main directions. First, addressing data movement and storage challenges is crucial for reducing \revb{energy} consumption and improving performance. Second, integrating and pipelining \revb{multiple} genome analysis steps \revb{using hardware-software co-design} can enhance efficiency by reducing \revd{both} \revb{useless computation and data movement}. Third, significant potential exists in enabling \revb{accurate and fast} real-time genome analysis by \revf{co-}developing efficient algorithms \revf{together} with specialized hardware, resulting in low-power, \revb{high-performance} and cost-effective \revb{(portable)} sequencing with \revc{low} latency.

\section*{Acknowledgments}

We thank the organizers of the DAC-60 conference for the invitation to contribute this invited paper and deliver an associated invited talk. We acknowledge many SAFARI Research Group Members who have contributed to some of the works described in this paper, especially Mohammed Alser and Damla Senol Cali, who have completed their PhD dissertations on the general topic of accelerating genome analysis. We thank all members of the SAFARI Research Group for the stimulating and scholarly intellectual environment they provide. We acknowledge the generous gift funding provided by our industrial partners (especially by Google, Huawei, Intel, Microsoft, VMware), which has been instrumental in enabling the decade+ long research we have been conducting on accelerating genome analysis. This work is also partially supported by the Semiconductor Research Corporation (SRC), the European Union’s Horizon programme for research and innovation [101047160 - BioPIM] and the Swiss National Science Foundation (SNSF) [200021\_213084].

% \bibliographystyle{unsrt}
% \setstretch{0.95}
\bibliographystyle{IEEEtran}
\bibliography{IEEEabrv, main, safari_bio, safari_pim}
\end{document}